\documentclass[twocolumn]{revtex4}
\usepackage{amsmath}
\usepackage{epsfig}
\makeatletter
\def\btt#1{\texttt{\@backslashchar#1}}%
\DeclareRobustCommand\bblash{\btt{\@backslashchar}}%
\makeatother

\begin{document}

\title{Object Picture of Quasinormal Modes for Stringy Black Holes}


\author{Ping Xi}
\author{Xin-zhou Li}\email{kychz@shnu.edu.cn}

\affiliation { Shanghai United Center for Astrophysics(SUCA),
Shanghai Normal University, 100 Guilin Road, Shanghai 200234,China
}%


\begin{abstract}
\section*{abstract}
We study the quasinormal modes (QNMs) for stringy black holes. By
using numerical calculation, the relations between the QNMs and the
parameters of black holes are minutely shown. For (1+1)-dimensional
stringy black hole, the real part of the quasinormal frequency
increases and the imaginary part of the quasinormal frequency
decreases as the mass of the black hole increases. Furthermore, the
dependence of the QNMs on the charge of the black hole and the
flatness parameter is also illustrated. For (1+3)-dimensional
stringy black hole, increasing either the event horizon or the
multipole index, the real part of the quasinormal frequency
decreases. The imaginary part of the quasinormal frequency increases
no matter whether the event horizon is increased or the multipole
index is decreased.
\end{abstract}


\maketitle

The elegant work on the quasinormal modes (QNMs) of a black hole was
carried out by Chandrasekhar [1], in which their role in the
response of the black hole to external perturbation. Since the
gravitational radiation excited by the black hole oscillation is
dominated by its QNMs, one can determine the parameters of a black
hole by analyzing the QNMs in its gravitational radiation. Thus,
besides their importance in the analysis of the stability of the
black hole, QNMs are important in the search for black holes and
their gravitational radiation. Many physicists believe that the
figure of QNMs is a unique fingerprint in directly identifying the
existence of a black hole. The QNMs of black holes in the framework
of general relativity have been studied widely [2]. On the other
hand, some solutions [3] can also be interpreted as black holes
whose parameters can be deduced from string theory by corresponding
compactifications. By studying the black hole in string theory,
researchers have successfully counted the black hole microstates
[4]. In previous work [5], we have studied the QNMs of stringy black
holes [6] by the semi-analytic method [7] and WKB method [8]. This
investigation has shown that the late-time gravitational oscillation
of the black hole under an external perturbation is dominated by
certain QNMs. The semi-analytic method is pioneered by Mashhoon and
his co-workers [7], in which the effective potential of the
Regge-Wheeler equation is replaced by a parameterized analytic
potential whose simply exact solutions are known. The parameters in
the potential are obtained by fitting the height, curvature and the
asymptotic value of the potentials. Therefore, the survey of how the
QNMs behaves is lacking for the various parameters of stringy black
holes using the above methods.

In this letter, we investigate in detail the relations between QNMs
of stringy black holes and their parameters by the numerical
calculation in null coordinates. Some results attained by this way
are supported by the semi-analytic results and WKB results. Most of
importance, we find more relations between the QNMs and the
parameters: the mass of the black hole, the flatness parameter and
the event horizon, respectively.

The exact solutions of string theory in the form of
(1+1)-dimensional black holes have attracted much interest [6].
These stringy black holes can be obtained by non-minimally coupling
the dilaton to gravitational system:
\begin{equation}
S=\int{d^{2}x\sqrt{G}e^{-2\phi}[R-4(\nabla{\phi})^2-c-{\frac{1}{4}}F^2]},
\end{equation}
where $\phi$ is the dilaton field, and $F_{\mu\nu}$ is the Maxwell
tensor. In the Schwarzschild-like gauge [5], we have the solutions
are
\begin{eqnarray}
&&\phi=\phi_0-\frac{Q}{2} r, F_{tr}=\sqrt{2} Q  q e^{-Qr},\nonumber\\
&&ds^{2}=f(r)dt^{2}-\frac{1}{f(r)}dr^{2},\\
&&f(r)=1-2me^{-Qr}+q^{2}e^{-2Qr},
\end{eqnarray}
where $m$ and $q$ can be considered as the mass and the charge of
the black hole, respectively; $Q$ and $\phi_{0}$ are integration
constants, and the asymptotic flatness condition requires $c=-Q^{2}$
and $Q>0$. When $m^{2}>q^{2}$, the event horizon of the black hole
lies at
\begin{equation}
r_{\pm}  = {\frac{{\ln \left( {m \pm \sqrt {m^{2} - q^{2}}}
\right)}}{{Q}}}.
\end{equation}

\noindent \indent Now, we consider concretely the behaviour of the
spacetime under the gravitational perturbation. The Regge-Wheeler
equation is [5]
\begin{equation}
{\frac{{d^{2}\psi (r^{ \ast} )}}{{dr^{ \ast 2}} }} + [\omega ^{2} -
V(r)]\psi (r^{ \ast }) = 0,
\end{equation}
where the tortoise coordinate $r^{\ast}$ is defined as
\begin{equation}
r^{\ast}=\int{\frac{dr}{f(r)}},
\end{equation}
and the effective potential
\begin{equation}
V(r^{\ast})={\frac{3}{4}}({\frac{df(r)}{f(r)dr^{\ast}}})^2-{\frac{1}{2}}{\frac{d}{dr^{^{\ast}}}(\frac{df(r)}{f(r)dr^{\ast}})}.
\end{equation}
\noindent \indent Obviously, the effective potential $V(r^{\ast})$
vanishes at $r\rightarrow\infty$, which corresponds to
$r^{\ast}\rightarrow\infty$, and is a finite value at the horizon or
$r\rightarrow$ 0 corresponding to the finite value of $r^{\ast}$ and
$r^{\ast}\rightarrow$ 0, respectively.\\
\noindent\indent We introduce the null coordinates $u = t - r^{\ast}
$ and $v = t + r^{\ast} $, Eq.(5) can be reduced to
\begin{equation}
-4\frac{\partial^{2}\psi}{\partial u \partial v}=V(r^{\ast})\psi.
\end{equation}
Equation (8) can be numerically integrated by the ordinary finite
element method. Using the Taylor expansion, we have
\begin{widetext}
\begin{equation}
\psi_{N}=\psi_{E}+\psi_{W}-\psi_{S} -\delta u \delta
v(\frac{v_{N}+v_{W}-u_{N}-u_{E}}{4})V\frac{\psi_{W}+\psi_{E}}{8}
+O(\Delta^{4}),
\end{equation}
\end{widetext}
where $N$, $W$, $E$ and $S$ are the points of a unit grid on the
$u-v$ plane which correspond to ($u+\Delta$, $v+\Delta$),
($u+\Delta$, $v$), ($u$, $v+\Delta$) and ($u$, $v$), and $\Delta$ is
the step length of the change of $u$ or $v$, i.e., $\Delta = \delta
u = \delta v$ [9]. Because the QNMs of stringy black holes is
insensitive to the initial conditions, we begin with a Gaussian
pulse of width $\sigma$ centered on $v_{c}$ on $u = u_{0}$ and set
the wave function to zero on $v = v_{0}$,
\begin{eqnarray}
&&\psi(u=u_{0},v)=\exp[-\frac{(v-v_{c})^{2}}{2\sigma^{2}}],\\
&&\psi(u,v=v_{0})=0.
\end{eqnarray}
Next, the point in the $u-v$ plane can be calculated by using
Eq.(8), successively. Finally, the values of $\psi(u_{max},v)$ are
extracted after the integration is completed where $u_{max}$
represents the maximum of $u$. Taking sufficiently large $u_{max}$
for the various $v$-value, we obtain a good approximation for the
wave function of the QNMs of the stringy black hole near the event
horizon.\\
 \noindent\indent One can see in Figs.1-3 that numerical
results of QNMs are shown for the various parameters of
(1+1)-dimensional stringy black hole. As a reminder, the oscillating
quasi-period and the damping time scale are shown in these figures.
In Fig. 1, we fixed $q=0.99$ and $Q=4$, and the wave functions vary
with the mass of the stringy black hole. Obviously, the oscillating
quasi-period decreases as the mass $m$ increases, but the damping
time scale slightly increases with the mass $m$. In other words, the
real part of the quasinormal frequency ($\omega_{R}$) increases and
the imaginary part of the quasinormal frequency ($\omega_{I}$)
decreases as the increase of the mass. In Fig. 2, we show the
relation between the quasinormal frequencies and the charge $q$. We
fixed $m=1$ and $Q=4$, and the wavefunctions vary with the charge of
the stringy black hole. Our numerical result is consistent with Ref.
[5], the oscillating quasi-period increases when the charge $q$
increases. In Fig. 3, we choose $m=1$ and $q=0.99$, and consider the
wavefunctions vary with the parameter $Q$. The oscillating frequency
and the damping time scale are both increasing as $Q$ increases.
\begin{figure}
\psfig{file=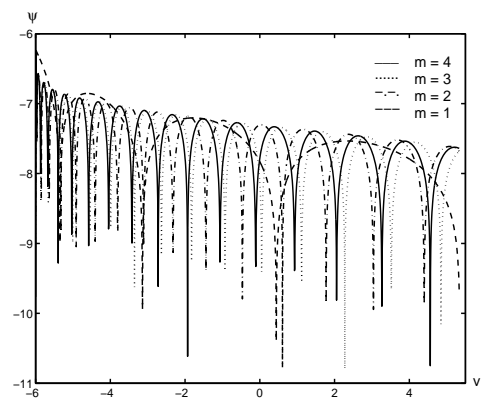,height=2.1in,width=2.5in} \caption{For
(1+1)-dimensional stringy black holes with $q=0.99$ and $Q=4$, the
wave functions are shown via different mass $m$.}
\end{figure}
\begin{figure}[!htbp]
\psfig{file=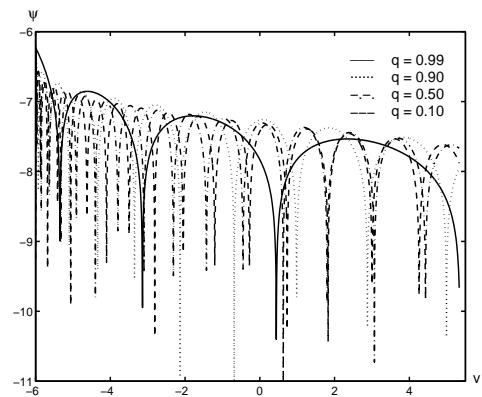,height=2.1in,width=2.5in}\caption{For
(1+1)-dimensional stringy black holes with $m=1$ and $Q=4$, the wave
functions are shown for different charge $q$.}
\end{figure}
\begin{figure}[!htbp]
\psfig{file=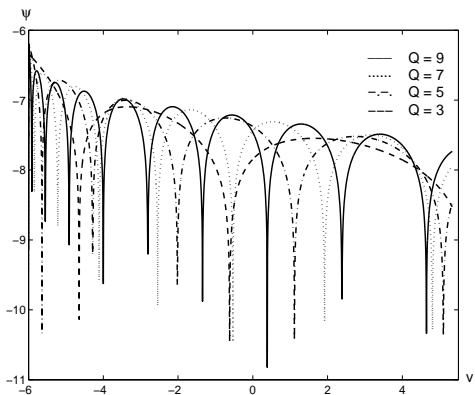,height=2.1in,width=2.5in}\caption{For
(1+1)-dimensional stringy black holes with $m=1$ and $q=0.99$, the
wave functions are shown for different constant $Q$. }
\end{figure}
In our previous work [5], we have obtained that the generic form of
the Regge-Wheeler equation corresponding to the external
perturbation for a (1+3)-dimensional stringy black hole [10] is
\begin{equation}
{\frac{{d^{2}\psi (r^{ \ast} )}}{{dr^{ \ast 2}} }} + [\omega ^{2} -
V(r)]\psi (r^{ \ast }) = 0,
\end{equation}
where the effective potential can be written as
\begin{eqnarray}
V(r)&=&{\frac{3r_{+}^{2}(1+6r_{+}^{2}+6r_{+}^{4})}{(r_{+}^{3}+r)^{4}}}
-{\frac{7r_{+}^{5}(1+3r_{+}^{2}+2r_{+}^{4})}{(r_{+}^{3}+r)^{5}}}\nonumber\\
&+&{\frac{4r_{+}^{8}(1+r_{+}^{2})^{2}}{(r_{+}^{3}+r)^{6}}}-{\frac{5r_{+}(1+2r_{+}^{2})}{(r_{+}^{3}+r)^{3}}}\nonumber\\
&+&{\frac{l^{2}+l-2\sqrt{r(r_{+}-r)}}{(r_{+}^{3}+r)^{4}}}+{\frac{2}{(r_{+}^{3}+r)^{2}}},
\end{eqnarray}
where $r_{+}$ is the event horizon, and generalized tortoise
coordinate takes the generic form
\begin{equation}
r^{\ast}=\int{\frac{(r_{+}^{3}+r)^{4}}{r(r-r_{+})}dr}.
\end{equation}
\noindent\indent Then, we calculate the QNMs for (1+3)-dimensional
stringy black hole using the finite element method. The numerical
results are shown in Figs. 4 and 5. In Fig. 4, we show how the QNMs
behave for various event horizons in the $l=20$ case. It is
interesting to note that the wavefunction experiences an increase of
the oscillating time and the damping time with an increase of the
event horizon. In Fig. 5, we discuss the relations between the QNMs
and the multipole index $l$ fixing $r_{+}=0.8$. It is clear that the
real part of the quasinormal frequency increases as the multipole
index $l$ increases, which is similar to the result in Ref. [5].
However, the imaginary part of the quasinormal frequency decreases
with $l$ increasing.
\begin{figure}[!hbp]
\psfig{file=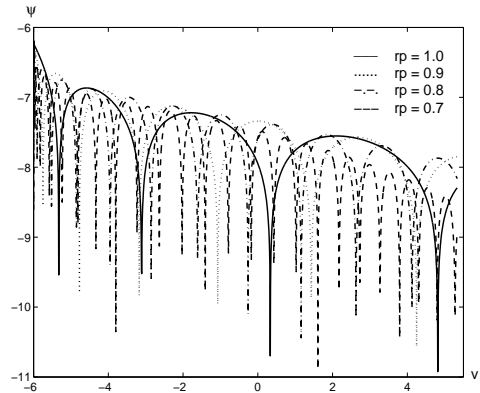,height=2.1in,width=2.5in}\caption{For
(1+3)-dimensional stringy black holes with $l=20$, the wave
functions are shown for different event horizons $r_{+}$.}
\end{figure}
\begin{figure}[!hbp]
\psfig{file=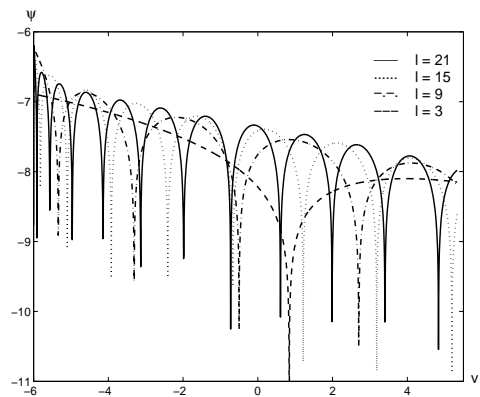,height=2.1in,width=2.5in}\caption{For
(1+3)-dimensional stringy black holes with $rp=0.8$, the wave
functions are shown for different multipole indexes $l$.}
\end{figure}

\section*{Acknowledgments}
This work is supported by the National Nature Science Foundation of
China under Grant No 10473007.

\end{document}